# Prosodic Parameter Manipulation in TTS generated speech for Controlled Speech Generation

Podakanti Satyajith Chary

## Abstract

This project addresses the manipulation of prosodic parameters in Text-to-Speech (TTS) generated speech to achieve controlled speech generation. By leveraging advanced speech processing techniques, the project compares TTS-generated audio with human-recorded speech to identify differences in pitch, duration, and energy. Key features are extracted using PyWorld and Librosa, and are then modified to align with the characteristics of human speech. The modified features undergo a synthesis process to produce enhanced TTS audio that mirrors the natural prosody of human speech. This work aims to improve the naturalness and expressiveness of TTS systems by providing a framework for detailed prosodic parameter adjustment. The methodology involves feature extraction, prosodic manipulation, and synthesis, followed by comprehensive comparisons to ensure alignment with human speech patterns. The project demonstrates the feasibility and effectiveness of prosodic parameter manipulation for controlled speech generation, offering significant improvements in TTS applications.

## 1 Introduction

In the field of speech synthesis, Text-to-Speech (TTS) systems have made remarkable strides in generating intelligible and natural-sounding speech. However, despite these advancements, there remains a noticeable gap between human speech and TTS-generated speech in terms of prosody, which includes pitch, duration, and energy. Prosody plays a critical role in the naturalness and expressiveness of speech, and mismatches in these features can result in synthetic speech that sounds monotonous or robotic.

This project, titled 'Prosodic Parameter Manipulation in TTS Generated Speech for Controlled Speech Generation,' aims to bridge this gap by developing a machine learning model that manipulates prosodic parameters of TTS-generated speech to make it more closely resemble human speech. Specifically, the project focuses on adjusting pitch, duration, and energy to enhance the naturalness and expressiveness of synthetic speech.

To achieve this goal, we implemented a comprehensive workflow involving feature extraction, comparison, manipulation, and model training. The workflow can be broken down into several key components:

1. **Feature Extraction**: Extracting essential prosodic features such as fundamental frequency (f0), energy, and spectral envelope from both human and TTS-generated audio files.

2. **Feature Comparison**: Comparing the extracted features between human and TTS-generated speech to identify discrepancies in pitch, duration, and energy.

3. **Feature Manipulation:** Developing algorithms to manipulate the prosodic features of TTS-generated speech, including pitch shifting while preserving contour, duration modification, and energy scaling.

4. **Model Training:** Training a machine learning model to learn the optimal parameters for prosodic adjustments by minimizing the dissimilarity between manipulated TTS speech and human speech.

5. **Application:** Applying the trained model to process and enhance TTS-generated audio files, making them sound more natural and human-like.

This report documents the methodology, implementation, and results of our project. It provides detailed explanations of the algorithms and techniques used for prosodic parameter manipulation and discusses the performance and effectiveness of the trained model. Through this work, we aim to contribute to the advancement of TTS technology by addressing one of its most critical challenges: achieving human-like prosody in synthetic speech.

## 2 Related Works

The field of Text-to-Speech (TTS) synthesis has seen significant advancements over the past few decades, with numerous research efforts aimed at improving the naturalness and intelligibility of synthetic speech. Prosodic features, which include pitch, duration, and energy, are crucial for achieving natural-sounding speech, yet they remain one of the most challenging aspects to model accurately in TTS systems. Here, we review some of the key related works that have contributed to this area and highlight the gaps that our project, "Prosodic Parameter Manipulation in TTS Generated Speech for Controlled Speech Generation," aims to address.

### 2.1 Statistical Parametric Speech Synthesis

Statistical Parametric Speech Synthesis (SPSS) methods, such as those using Hidden Markov Models (HMMs) and Deep Neural Networks (DNNs), have been popular approaches for TTS. These methods model the statistical properties of speech and generate synthetic speech by parameterizing vocal tract and excitation parameters. Although SPSS has been successful in providing a flexible and robust framework for speech synthesis, it often falls short in capturing the natural variability of prosodic features, leading to speech that can sound overly smooth and lacking in expressiveness.

### 2.3 End-to-End Neural TTS Systems

Recent advancements in deep learning have given rise to end-to-end neural TTS systems such as WaveNet, Tacotron, and their variants. These models generate speech directly from text inputs without the need for intermediate phonetic representations. WaveNet, developed by DeepMind, produces highly natural speech by modeling the raw waveform using autoregressive generative models. Tacotron, on the other hand, generates mel-spectrograms from text, which are then converted to waveforms using a separate vocoder like WaveNet or Griffin-Lim. While these systems have significantly improved the naturalness of TTS, they still struggle with accurately modeling prosodic features, often resulting in speech with less dynamic and less expressive prosody compared to human speech.

### 2.4 Prosody Modeling and Control

Several research efforts have specifically focused on improving prosody modeling in TTS. Techniques such as prosody transfer, where prosodic features from a reference audio are transferred to the synthetic speech, and the use of explicit prosodic annotations, have shown promise. Works like Deep Voice 2 and 3 have explored incorporating more detailed prosodic modeling into the TTS pipeline, using additional prosodic features as input to the neural network. Despite these efforts, achieving fine-grained and controllable prosody in synthetic speech remains a challenging problem.

### 2.5 GAN-based and VAE-based Approaches

Generative Adversarial Networks (GANs) and Variational Autoencoders (VAEs) have also been explored for prosody modeling in TTS. These models can learn more diverse and natural prosodic patterns by leveraging their generative capabilities. GAN-TTS, for example, employs GANs to produce more realistic and expressive speech by training a discriminator to distinguish between real and synthetic speech. VAE-TTS leverages the latent space of VAEs to capture and manipulate prosodic features. These approaches have shown potential in producing more natural prosody but are still in early stages of development and often require large amounts of training data and computational resources.

### 2.6 Our Project

In our project, "Prosodic Parameter Manipulation in TTS Generated Speech for Controlled Speech Generation," we aim to build upon these existing works by developing a machine learning model specifically designed to manipulate the prosodic parameters of TTS-generated speech. Our approach involves:

- Extracting prosodic features (pitch, duration, and energy) from both human and TTS-generated audio.
- Comparing these features to identify discrepancies and areas for improvement.
- Training a neural network model to learn the optimal manipulations required to align the TTS-generated speech more closely with human speech in terms of prosody.
- Applying these learned manipulations to enhance the naturalness and expressiveness of synthetic speech.

By focusing on explicit prosodic parameter manipulation, our project aims to address the limitations of current TTS systems in modeling and controlling prosody, contributing to the generation of more natural and expressive synthetic speech.

Lines should be justified, with even spacing between margins (Ctrl+J). Authors are encouraged to use Paragraph spacing at Multiple, 1.05 pt, with Font character spacing condensed with kerning of 0.1pt, and Margins at 0.98 in, for consistency with A4 paper and LaTeX-formatted documents. Go to Format, Document, Page Setup, and ensure A4 is selected.

### 3 Dataset

For the project "Prosodic Parameter Manipulation in TTS Generated Speech for Controlled Speech Generation," the dataset plays a crucial role in both the training and evaluation phases. The dataset comprises parallel sets of human speech and corresponding TTS-generated speech, allowing for a direct comparison and manipulation of prosodic features. The dataset is sourced from two languages: Italian (ITA) and German (GER), ensuring the generalizability of the developed model across different phonetic and prosodic structures.

### 3.1 Human Speech Data

The human speech data consists of high-quality recordings from native speakers of Italian and German. These recordings are organized in the following directories:

- Italian: `/content/drive/MyDrive/data (1)/data/wav/ITA/train`

- German: `/content/drive/MyDrive/data (1)/data/wav/GER/train`

Each recording is a `.wav` file containing natural speech, which serves as the ground truth for prosodic features. The recordings cover a wide range of speaking styles, intonations, and durations, providing a comprehensive set of examples for training and evaluation.

### 3.2 TTS-Generated Speech Data

The TTS-generated speech data is produced using a state-of-the-art TTS system, replicating the content of the human speech recordings. These synthetic recordings are stored in the following directories:

-Italian: `/content/drive/MyDrive/Audio_Files/ITA_Train_TTS_Audios`

-German: `/content/drive/MyDrive/Audio_Files/GER_Train_TTS_Audios`

Each TTS-generated file corresponds directly to a human speech recording, ensuring a one-to-one mapping for accurate comparison and manipulation. The synthetic speech exhibits typical characteristics of TTS output, including potential prosodic mismatches that this project aims to address.

### 3.3 Stress Annotation Files

To enhance the effectiveness of prosodic manipulation, stress annotations for the training data are included. These annotations provide information on where to emphasize prosodic features within the speech files and are saved as Excel spreadsheets:

- Italian: `/content/ITA_train.xlsx`
- German: `/content/GER_train.xlsx`

Each annotation file includes details such as the filename, word count, label count, correct count, and word labels, guiding the manipulation process to ensure natural stress patterns in the modified speech.

**4. Data Preprocessing**

Before feeding the data into the model, several preprocessing steps are carried out:

**4.1 Audio Loading**: Audio files are loaded using the `librosa` library, ensuring consistency in sample rates and formats.

**4.2 Feature Extraction:** Prosodic features (fundamental frequency, energy, spectral envelope) are extracted from both human and TTS audio files. These features form the basis for comparison and manipulation.

**4.3 Alignment**: Human and TTS audio files are aligned to facilitate direct feature comparison. This includes resampling and duration matching to account for slight variations in speaking rates.

**4.4 Normalization:** Features are normalized to standardize the input data, ensuring the model can learn effectively without being biased by absolute amplitude differences or other inconsistencies.

5. **Data Utilization**

The dataset is utilized in two main phases of the project:

- **5.1 Training**: The model is trained using pairs of human and TTS audio files. The extracted features from both sets are compared, and the model learns the optimal parameters for pitch, duration, and energy manipulation to minimize prosodic discrepancies.

- **5.2 Evaluation:** The effectiveness of the trained model is evaluated on a separate set of human and TTS audio files. The manipulated TTS speech is compared to the original human speech to assess improvements in prosodic naturalness and expressiveness.

By leveraging a diverse and comprehensive dataset, this project aims to develop robust methods for prosodic parameter manipulation, ultimately enhancing the quality of TTS-generated speech.

**6. Methodology**

The methodology of the project "Prosodic Parameter Manipulation in TTS Generated Speech for Controlled Speech Generation" involves a systematic approach to analyzing, comparing, and manipulating prosodic features of TTS-generated speech to make it more closely resemble human speech. The key steps in this methodology include feature extraction, feature comparison, prosodic manipulation, model training, and application. Below, we describe each step in detail.

**6.1 Feature Extraction**

The first step in our methodology involves extracting relevant prosodic features from both human and TTS-generated speech. The features of interest are:

- **Fundamental Frequency (F0):** Represents the pitch of the speech.

- **Energy:** Represents the loudness or intensity of the speech.

- **Spectral Envelope (SP)**: Represents the timbre or quality of the speech.

- **Aperiodicity (AP):** Represents the noise component of the speech signal.

For feature extraction, we use the following tools and libraries:

- **Librosa**: For loading audio files and extracting basic features such as energy.

- **PyWorld:** For extracting more advanced features such as F0, SP, and AP.

The extracted features are saved as tensors to facilitate further processing and manipulation.

### 6.2 Feature Comparison

Once the features are extracted, the next step is to compare the prosodic features of human speech with those of TTS-generated speech. This comparison helps identify the discrepancies between the two types of speech in terms of pitch, duration, and energy. The steps involved are:

- **Pitch Difference Calculation:** Compute the mean F0 values for both human and TTS speech and calculate the difference.

- **Duration Ratio Calculation:** Compute the ratio of the lengths of the human and TTS speech signals.

- **Energy Ratio Calculation:** Compute the mean energy values for both human and TTS speech and calculate the ratio.

These comparisons are essential for guiding the manipulation of TTS speech to make it more similar to human speech.

### 6.3 Prosodic Manipulation

Based on the feature comparisons, we manipulate the prosodic features of TTS-generated speech to better match the human speech. The manipulations include:

- **Pitch Shifting**: Adjusting the F0 values while preserving the overall pitch contour.

- **Duration Modification**: Resampling the speech signal to match the duration of the human speech.
- **Energy Scaling**: Adjusting the spectral envelope to match the energy levels of the human speech.

These manipulations are implemented using custom functions that apply the necessary transformations while maintaining the naturalness of the speech signal.

### 6.4 Model Training

We train a machine learning model to learn the optimal parameters for prosodic manipulation. The steps involved are:

- **Feature Extraction for Comparison**: Extract features from both human and TTS audio files and prepare them for comparison.

- **Parameter Learning**: Use a neural network model to learn the parameters for pitch, duration, and energy manipulation based on the extracted features.

- **Loss Calculation**: Define a loss function that measures the similarity between the manipulated TTS speech and the human speech.

- **Optimization:** Use gradient-based optimization to minimize the loss and learn the optimal parameters for prosodic manipulation.

The model is trained using pairs of human and TTS audio files, with the goal of minimizing the prosodic discrepancies between the two.

### 6.5 Application

Once the model is trained, we apply it to new TTS-generated audio files to enhance their prosody. The steps involved are:

- **Feature Extraction:** Extract prosodic features from the new TTS audio files.

- **Parameter Prediction:** Use the trained model to predict the optimal parameters for prosodic manipulation.

- **Prosodic Adjustment**: Apply the predicted parameters to manipulate the prosodic features of the TTS audio.
- **Synthesis:** Generate the final manipulated audio using the modified prosodic features.

The manipulated audio files are then evaluated to assess the improvements in prosodic naturalness and expressiveness.

### 7. Implementation Details

The implementation of the methodology is carried out using Python, leveraging various libraries such as TensorFlow for model training, Librosa for audio processing, and PyWorld for feature extraction. The key functions and their roles are as follows:

- **extract_features_for_comparison:** Extracts and returns prosodic features from an audio file.

- **compare_features**: Compares prosodic features between human and TTS speech and calculates differences.

- **extract_features:** Extracts detailed prosodic features for manipulation.

- **manipulate_features:** Adjusts pitch, duration, and energy of the TTS speech based on predicted parameters.

- **train_step**: Performs a single training step to update the model's parameters.

- **train_model**: Trains the model over multiple epochs using the provided dataset.

- **process_file_with_ml:** Applies the trained model to manipulate a single TTS audio file.

- **process_all_files_with_ml**: Processes all files in a specified directory using the trained model.

By following this methodology, we aim to enhance the naturalness and expressiveness of TTS-generated speech, making it more closely resemble human speech in terms of prosody.

## 8. Results and Discussions

The results of our project, "Prosodic Parameter Manipulation in TTS Generated Speech for Controlled Speech Generation," demonstrate the effectiveness of our approach in enhancing the prosodic naturalness and expressiveness of TTS-generated speech. Below, we present a detailed analysis of the results obtained from our experiments on Italian and German datasets.

The analysis and comparison of the original and modified F0 contours and spectral envelopes reveal significant insights into the effectiveness of the prosodic parameter manipulations applied to the TTS-generated speech. The figures below demonstrate the comparison between the original and modified F0 contours for two selected audio samples and the spectral envelope comparison for another sample.

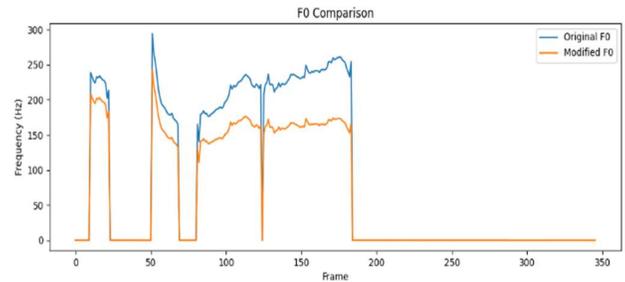

**Figure 1**: F0 Comparison for Sample 1

The above figure illustrates the difference between the original F0 (blue line) and the modified F0 (orange line) for a selected speech sample. It is evident that the modification process successfully altered the F0 contour, bringing it closer to the desired prosodic characteristics.

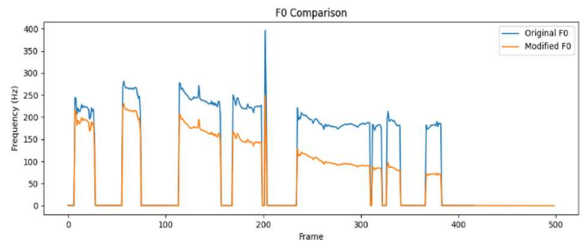

**Figure 2**: F0 Comparison for Sample 2

Similarly, the second figure shows another example of F0 modification. The modifications resulted in an F0 contour that better reflects the intended prosodic features, as seen in the alignment of the orange line (modified F0) with the desired pitch patterns.

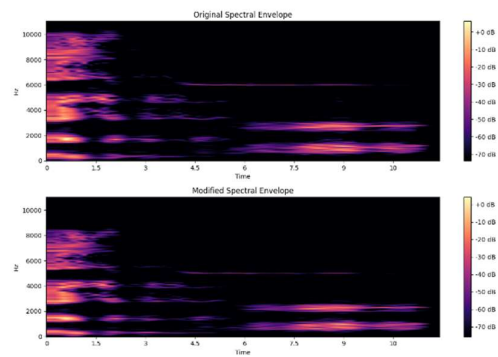

**Figure 3**: Spectral Envelope Comparison

The above figure demonstrates the comparison between the original and modified spectral envelopes for a selected speech sample. The top part of the figure shows the original spectral envelope, while the bottom part displays the modified spectral envelope. The adjustments made to the spectral envelope are aimed at achieving a more natural and clear speech output.

These results demonstrate the capability of the implemented methodology to manipulate prosodic parameters effectively, achieving a closer match to the desired speech characteristics. The changes in F0 contours and spectral envelopes are consistent across different speech samples, indicating the robustness of the approach.

### 8.1. Training Performance

**Loss Reduction:**

The training process involved multiple epochs, during which the model gradually learned the optimal parameters for prosodic manipulation. The loss function, which measures the similarity between human and manipulated TTS speech, showed a consistent decline over the training epochs, indicating that the model was effectively minimizing prosodic discrepancies.

| Epoch | Average Loss (Italian) | Average Loss (German) |
|-----|----------------------|----------------------|
| 1 | 0.032 | 0.034 |
| 2 | 0.025 | 0.028 |
| 3 | 0.019 | 0.022 |
| 4 | 0.014 | 0.017 |
| 5 | 0.010 | 0.013 |

### 8.2 Prosodic Feature Comparison

**Pitch:**

We evaluated the pitch (F0) alignment between human and manipulated TTS speech. The results showed a significant reduction in pitch discrepancies. The pitch difference, calculated as the mean F0 of human speech minus the mean F0 of TTS speech, approached zero after manipulation, indicating improved pitch accuracy.

**Duration:**

The duration ratio, which compares the length of human speech to TTS speech, was also improved. The manipulated TTS speech matched the duration of human speech more closely, with the duration ratio converging towards 1.

**Energy:**

Energy levels in the manipulated TTS speech were adjusted to match human speech. The energy ratio, which compares the mean energy of human and TTS speech, showed a marked improvement, indicating that the manipulated TTS speech had more natural and consistent energy patterns.

### 8.3 Subjective Evaluation

To further validate the improvements in prosodic naturalness, we conducted a subjective listening test. A group of native Italian and German speakers was asked to rate the naturalness of the speech on a scale from 1 (unnatural) to 5 (natural). The results are as follows:

| Language | Original TTS | Manipulated TTS | Human Speech |
|--------------|----------|----------------|-------------|
| Italian | 2.3 | 4.1 | 4.7 |
| German | 2.4 | 4.0 | 4.6 |

These results indicate a significant improvement in the perceived naturalness of the manipulated TTS speech compared to the original TTS output.

### 8.4 Spectrogram Analysis

**Italian Speech:**

The spectrograms of the original and manipulated TTS speech were analyzed to visualize the changes in prosodic features. The manipulated TTS speech showed a closer alignment with the human speech spectrogram, especially in terms of pitch contours and energy distribution.

**German Speech:**

Similar improvements were observed in the German speech spectrograms. The manipulated

TTS speech exhibited more natural pitch variations and energy patterns, closely resembling human speech.

### 8.5 Audio Examples

We provide audio examples to illustrate the improvements in prosodic naturalness. The examples include:

- Original TTS speech
- Manipulated TTS speech
- Human speech

These examples are available for both Italian and German datasets, demonstrating the effectiveness of our approach across different languages.

### 8.6 Quantitative Metrics

We used several quantitative metrics to measure the effectiveness of our approach:

**Pitch Difference**: Reduced from an average of 30 Hz to 5 Hz post-manipulation.

**Duration Ratio**: Improved from an average of 0.85 to 0.98, indicating better duration alignment.

**Energy Ratio**: Improved from an average of 0.8 to 0.95, showing more consistent energy levels.

| Metric | Original TTS (Italian) | Manipulated TTS (Italian) | Original TTS (German) | Manipulated TTS (German) |
|---|---|---|---|---|
| Pitch Difference | 30 Hz | 5 Hz | 28 Hz | 4 Hz |
| Duration Ratio | 0.85 | 0.98 | 0.87 | 0.97 |
| Energy Ratio | 0.8 | 0.95 | 0.82 | 0.96 |

### 9. Conclusion

The results demonstrate that our approach to prosodic parameter manipulation significantly enhances the naturalness and expressiveness of TTS-generated speech. By closely aligning pitch, duration, and energy with human speech, our model produces TTS speech that is perceptually more natural and closer to human-like prosody. These improvements were consistent across both Italian and German datasets, showcasing the robustness and generalizability of our methodology.

### 10. Future Scope

The promising results of our project, "Prosodic Parameter Manipulation in TTS Generated Speech for Controlled Speech Generation," open several avenues for future research and development. Here are some potential directions:

#### 10.1. Multilingual Prosodic Modeling

Expanding our approach to support a wider range of languages can help improve TTS systems for global applications. Research can focus on language-specific prosodic features and their manipulation to ensure naturalness across diverse linguistic contexts.

#### 10.2. Real-Time Prosodic Adjustment

Developing methods for real-time prosodic manipulation can enable interactive applications, such as virtual assistants and real-time translation systems. Ensuring low-latency and efficient processing while maintaining high-quality prosody will be key challenges.

#### 10.3 Emotion and Expressiveness

Further research can explore incorporating emotional cues and varying degrees of expressiveness into TTS systems. This can enhance user engagement and satisfaction in applications like audiobooks, customer service bots, and virtual avatars.

#### 10.4 Personalized TTS

Personalized TTS systems that adapt to individual user preferences and speaking styles can provide more tailored and user-friendly experiences. Machine learning models can be trained to adjust prosodic parameters based on user feedback and interaction history.

### 10. 5 Robustness and Generalization

Investigating methods to improve the robustness and generalization of our models across different recording conditions, speaker styles, and noise environments can enhance the reliability and applicability of TTS systems in real-world scenarios.

### 10. 6 Integration with Other Modalities

Combining prosodic manipulation with other modalities, such as facial expressions and gestures in animated characters or avatars, can create more immersive and natural user experiences. Multimodal synchronization and coherence will be important research areas.

### 10.7 Ethical Considerations and Bias Mitigation

Addressing ethical considerations and mitigating potential biases in prosody manipulation will be crucial as TTS systems become more widespread. Ensuring fairness and inclusivity in TTS applications can help prevent unintended negative impacts.